\documentclass[aps,pra,twocolumn,superscriptaddress,groupedaddress,showpacs]{revtex4}

\usepackage{natbib}
\usepackage{amssymb,amsmath,latexsym}
\usepackage{amsthm}
\usepackage{hyperref}
\usepackage{bm}
\usepackage{microtype}
\usepackage{url}
\usepackage{graphicx}
\usepackage{enumitem}

\usepackage[lined,ruled,vlined]{algorithm2e}

\usepackage{xcolor}
\theoremstyle{plain}

\theoremstyle{definition}

\theoremstyle{remark}

\newcommand{\ket}[1]{\left| #1 \right\rangle}

\usepackage{xargs}
\usepackage[colorinlistoftodos,prependcaption]{todonotes}
\newcommandx{\unsure}[2][1=]{\todo[linecolor=red,backgroundcolor=red!25,bordercolor=red,#1]{#2}}
\newcommandx{\change}[2][1=]{\todo[linecolor=blue,backgroundcolor=blue!25,bordercolor=blue,#1]{#2}}
\newcommandx{\info}[2][1=]{\todo[linecolor=green,backgroundcolor=green!25,bordercolor=green,#1]{#2}}
\newcommandx{\improvement}[2][1=]{\todo[linecolor=purple,backgroundcolor=purple!25,bordercolor=purple,#1]{#2}}
\newcommandx{\thiswillnotshow}[2][1=]{\todo[disable,#1]{#2}}

\newcommand{\Complex}{\mathbb{C}}
\newcommand{\id}{\mathsf{id}}

\usepackage{listings} 

\definecolor{lightgray}{gray}{0.9}
\newcommand{\inlinecode}[1]{\colorbox{lightgray}{\lstinline[language=Python]$#1$}}

\begin{document}

\title{SimulaQron - A simulator for developing quantum internet software}
\author{Axel Dahlberg}
\affiliation{QuTech, Lorentzweg 1, 2628 CJ Delft, Netherlands}
\author{Stephanie Wehner}
\affiliation{QuTech, Lorentzweg 1, 2628 CJ Delft, Netherlands}
\email[]{s.d.c.wehner@tudelft.nl}

\begin{abstract}
We introduce a simulator of a quantum internet with the specific goal to support software development. 
A quantum internet consists of local quantum processors, which are interconnected by quantum communication channels that enable the transmission of qubits
between the different processors. While many simulators exist for local quantum processors, there is presently no simulator for a quantum internet tailored towards software development. Quantum internet protocols require both classical as well as quantum information to be exchanged between the network nodes, next to the execution of gates and measurements on a local quantum processor. This requires quantum internet software to integrate classical communication programming practises with novel quantum ones. 

SimulaQron is built to enable application development and explore software engineering practises for a quantum internet. 
SimulaQron can be run on one or more classical computers to simulate local quantum processors, which are transparently connected in the background to
enable the transmission of qubits or the generation of entanglement between remote processors. Application software can access the simulated local quantum processors to execute local quantum instructions and measurements, but also to transmit qubits to remote nodes in the network. 
SimulaQron features a modular design that performs a distributed simulation based on any existing simulation of a quantum computer capable of integrating with Python. Programming libraries for Python and C are provided to facilitate application development. 
\end{abstract}

\maketitle

\section{Introduction}\label{sec:intro}
A quantum internet enables quantum communication between remote quantum processors in order to solve problems that are infeasible classically.
Many applications of a quantum internet are already known, the most famous of which is quantum key distribution (QKD)~\cite{bb84, e91} which allows
two network nodes to establish an encryption key.
Examples of other applications are secure identification~\cite{secureID} and other two-party cryptographic tasks~\cite{Kaniewski2016},
clock synchronization~\cite{Jozsa2000}, secure delegated quantum computation~\cite{Childs2001}, and even extending the baseline of telescopes~\cite{telescope}.
For many of these applications, only relatively simple quantum processors capable of operating on a handful of qubits
are required, as they draw their power from quantum entanglement.
Entanglement can be realized using already one qubit at each end point, and its capabilities cannot be replicated using classical communication.
In the case of QKD, for example, quantum processors capable only of preparing and measuring single qubits can already be sufficient~\cite{bb84}.

The first quantum networks that connect remote quantum processors capable of operating on several qubits each are expected to be deployed within the coming years.
End-nodes, holding such quantum processors and on which applications run, will be able to send qubits and generate entanglement between each other using the network.
Depending on the distance between the end-nodes, different ways to realize the quantum communication can be used, including the use of
 quantum repeaters~\cite{Muralidharan2016,VanMeter2014,Sangouard2011,Azuma2015,Munro2015}.
In order to execute arbitrary quantum internet applications on these networks, it is essential to create a development framework in which software for these applications, running on the end-nodes, can be written and debugged.
Writing software for a quantum internet shares some similarities with programming a quantum computer, but in addition poses new challenges.
Similar to programming a quantum computer, we wish to execute quantum gates and measurements on each local quantum processor.
Techniques for optimizing such gates and mapping them to the underlying 
hardware can be borrowed from quantum computing efforts, and are hence not the subject of this work.
What differentiates programming a quantum network is the need for a close integration between classical and quantum messages exchanged during the course of the protocol.
The need for such integration arises at several layers of abstraction, and poses significant design challenges.
On a lower level, a quantum network stack is needed to create and track entanglement in the network, which requires classical control messages to be exchanged.
On a higher level, applications which desire to create and use such entanglement do themselves exchange classical messages during the course of a protocol, requiring standard classical network programming techniques to integrate with quantum ones.
At present, a full quantum network stack is missing, and no software development framework for writing quantum network applications exists.

\begin{figure}[!h]
    \centering
    \includegraphics[width=0.47\textwidth]{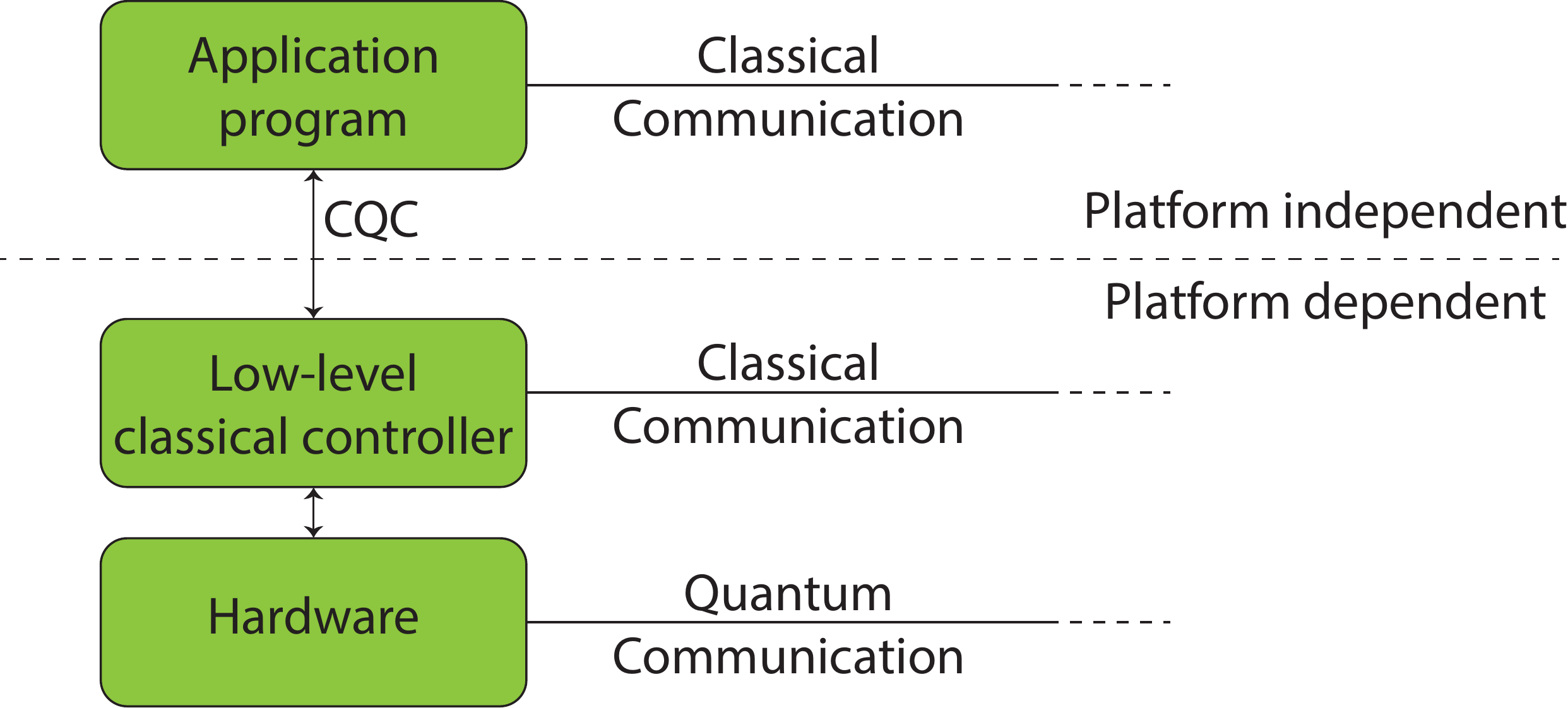}
    \caption{High-level schematic of the quantum network system of one node in the network. At the lowest level lies the quantum hardware, which in the most general case consists of a quantum processor capable of storing and manipulating qubits. The quantum processor has an optical interface over which it can generate entanglement with neighbouring network nodes, and send/receive qubits. Sending and receiving qubits may - depending on the implementation - be realized by first generating
        entanglement, followed by quantum teleportation. At present, quantum network nodes that support local quantum processing as well as the generation of long-lived entanglement are NV centers in diamond~\cite{Humphreys2017} and Ion Traps~\cite{Inlek2017}. The quantum hardware is controlled by a necessarily platform dependent control system that may possess a classical communication interface to neighbouring nodes, for example to communicate the successful generation of heralded entanglement~\cite{Humphreys2017}. Together the quantum hardware and low-level control system form the platform dependent quantum processing system.
SimulaQron provides a stand-in for such platform dependent quantum processing systems, and the quantum communication between them.
The platform dependent system provides a universal interface, which we call CQC (Classical Quantum Combiner). CQC can be understood as an extended instruction set which - next to supporting ``standard'' quantum instructions such as performing gates or measurements - features special features tailored to a quantum network. This includes, for example, commands to produce entanglement or transmit qubits. 
Applications can be realized in the platform independent part of the quantum internet system by sending the appropriate instructions using CQC to the underlying quantum processing system.
For simplicity, SimulaQron allows for direct communication between any two nodes in the network.
However, a different topologically can be realized by using SimulaQron in a restricted fashion.
Quantum network applications generally require the exchange of classical communication, and the integration between such classical communication and the use of the local quantum processing system is an integral aspect of application software
development. In practice, this classical communication may be solved transparently using the same physical medium as used for the low level control, but this is not a requirement.}
    \label{fig:cqc_simple}
\end{figure}

\subsection{What SimulaQron does}

SimulaQron is a simulator providing a tool for software development for a quantum internet, freely available online~\cite{simulaqron}.
Specifically, SimulaQron
simulates several quantum processors held by the end-nodes of the network, connected by a simulated quantum communication channels. This allows the simulation of single quantum networks, as well
as inter-connected quantum networks forming a quantum internet. 
The quantum communication channels between the end-nodes in a real implementation of the network, can be realized in different ways, for example using quantum repeaters.
The simulation of quantum repeaters and their performance is an important aspect of developing a quantum network.
Numerous simulations of quantum repeaters have been conducted such as for example~\cite{2058-9565-3-3-034002,Krovi2016,Muralidharan2014}, the objective of SimulaQron however is a rather different one in that it aims provide a platform for application development to software engineers.
SimulaQron can be used to develop the software running at the end-nodes together with classical communication between these.
Figure~\ref{fig:cqc_simple} provides a high level schematic of such a quantum internet system, where SimulaQron should be understood
as a stand-in for the quantum processing system (platform dependent) in order to enable platform independent software development to proceed
without access to quantum hardware. 

\begin{figure}[h]
    \centering
    \includegraphics[width=0.47\textwidth]{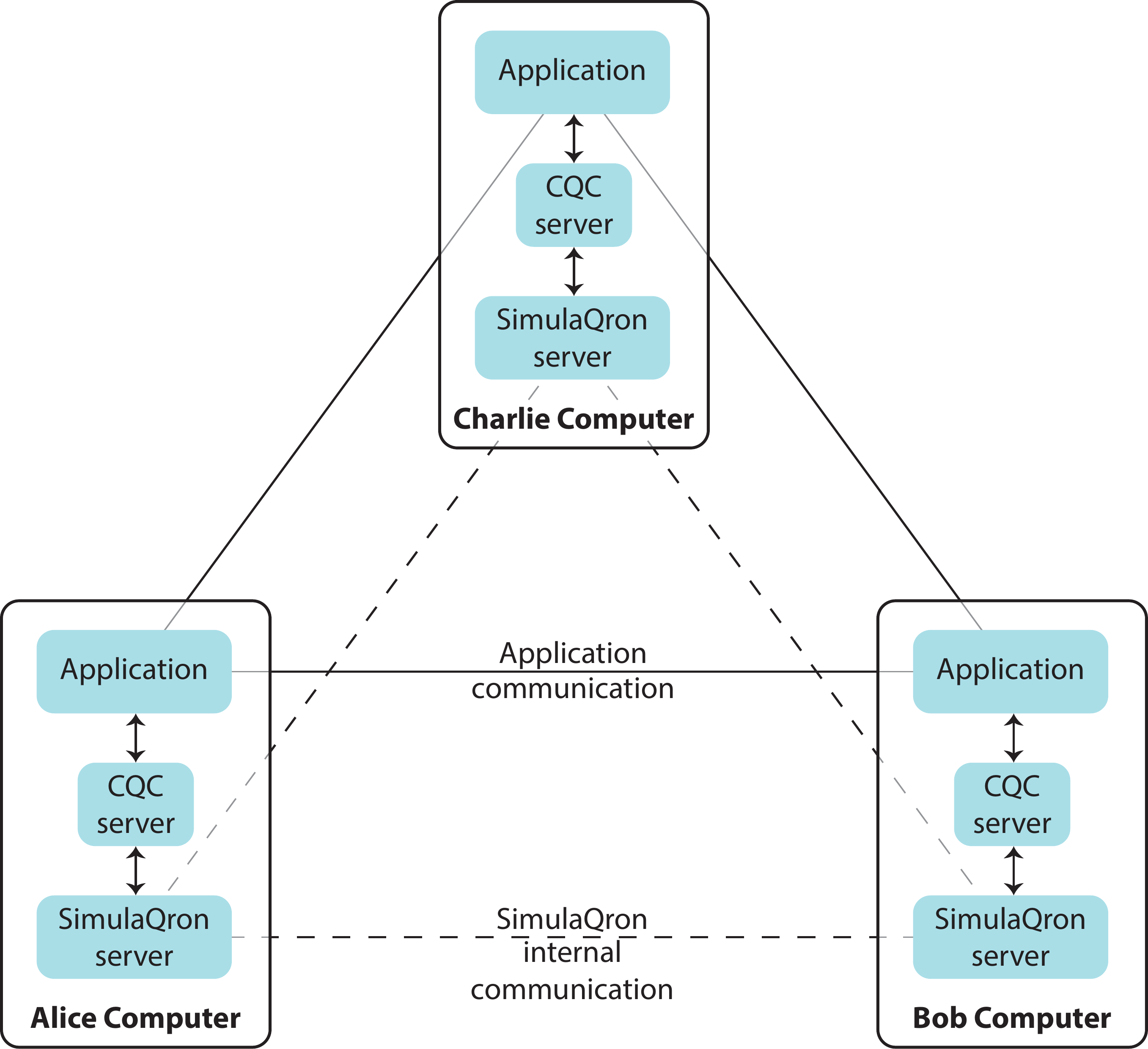}
    \caption{A schematic overview of the communication in a quantum network simulated by SimulaQron.  The simulation of the quantum hardware at each node is handled by the SimulaQron backend server.  Communication between the SimulaQron servers is needed to simulate the network, for example to simulate entanglement.  Opting for this method enables a distributed simulation, i.e. the computers in the figure can be physically different computers.  The CQC servers provide the CQC interface (see
    Figure~\ref{fig:cqc_simple}) to the applications running on the network and the simulated platform dependent quantum processing system (see also Figure~\ref{fig:cqc_schematic}). Internally the CQC servers establish a connection to the backend. The choice to use two distinct servers is motivated by modularity, in that the same CQC servers can in principle give access to multiple backends (see also Section~\ref{sec:doesNot}). Finally the applications can communicate classically, as they would in a real implementation of a quantum network.} \label{fig:servers}
\end{figure}

SimulaQron precisely mimicks a real network, and allows each simulated processor to run on a different classical computer. Each local processing system 
supports the execution of local quantum gates and measurements, but also network specific commands, for example to generate entanglement.
SimulaQron transparently simulates the exchange of qubits and the creation of entanglement between remote processors in the background, making this functionality available to applications. This is achieved by classical communication between the computers hosting the simulated quantum processors, as depicted in the schematic of Figure~\ref{fig:servers}. 

To perform the simulation of local quantum processors itself, SimulaQron uses existing simulators of quantum processors. What's more, SimulaQron's modular structure does in principle allow any simulation of a quantum computer accessible via Python to serve as a backend.
The key novelty in SimulaQron is to leverage such 
backends into a distributed simulation that maps locally simulated entangled qubits to remote network nodes, in order to simulate the availability of entanglement between distant quantum computers.
We emphasize that applications using SimulaQron's simulated entanglement, evidently do not provide the security guarantees afforded 
by real entanglement.
The objective is instead to use SimulaQron as a development platform to write software realizing these applications which can later run on real quantum hardware and use real entanglement in order to achieve these guarantees.
SimulaQron can be used as a tool for software development in all areas ranging from the implementation of the actual applications, the development of application level abstractions and programming libraries, to exploring and implementing a quantum network stack. We remark that there is presently no such stack. 

\subsection{What SimulaQron does not}\label{sec:doesNot}
We emphasize that SimulaQron is written as a tool for software engineers, with the objective of allowing software engineering efforts for a quantum internet to proceed.
Its goal is thus to allow developers to write software that can later run with no or minimal modification on real quantum internet hardware. 

For quantum experts, we remark that SimulaQron does not aim to achieve an efficient simulation of a large scale quantum internet in order to test quantum repeater schemes, error correcting codes, or study the effects of noise on distant qubits. Evidently, given that SimulaQron can use any local simulation of a quantum processor in Python 
as a backend, it is straightforward to let it use a backend simulating noisy qubits. Noise can also be added manually by the application, for example, by 
probabilistically applying Pauli gates to the qubits during the protocol. We remark however that such adhoc additions of noise do not allow us to make general 
and accurate statements about the performance of quantum network applications in the presence of noise. Noise in quantum devices is highly time dependent, and hence the amount of noise an application experiences is highly dependent on time delays - for example, how long it takes classical and quantum data to travel from one node to another. If SimulaQron would simply be programmed to apply a time dependent noise based on the message delays during simulation, then this would only be meaningful if these delays mirrored the exact time delays in the real network. What's more, even for executing local gates alone, the execution time of the simulation on the classical computer does not provide reliable timings. 

In order to explore the precise effect of noise in a quantum network, it is essential to be able to model time very precisely. This can be achieved using 
a technique called discrete event simulation that is well known from classical networking. This is subject of a separate simulation platform (NetSquid~\cite{netsquid}), which 
performs a discrete event simulation of a large scale quantum internet capable of precisely modelling timing delays and hence study the effects of delays in quantum communication, as well as classical control communication on the performance of quantum protocols and a quantum network stack. NetSquid is not yet publicly 
available and differs from SimulaQron in several key aspects apart from its ability to model time: first, it is a monolithic simulation and does not provide a real simulation network on different computers as SimulaQron does. 
As such, it does not mirror software engineering practices in networked programming. 
Second, it does not provide a real time interactive experience of working on a quantum network as provided by SimulaQron. 
We remark that NetSquid will feature a network emulation mode, in which applications written for SimulaQron can communicate via CQC with a - purely local - simulated quantum network using NetSquid in which time can be tightly controlled. This way their performance can be tested under controlled timing conditions, and realistic time dependent noise models of different hardware platforms.  

Also for software engineers, we remark that there are some things which are purposefully not handled by SimulaQron.
This includes, for example, the management and tracking of entanglement required by a quantum network stack, which may on the other hand be explored and implemented using SimulaQron. No quantum network stack, nor protocols for managing and tracking entanglement are presently known, and are still undergoing development~\cite{forwardCitation}.
Importantly, we also note that for the same reason no efforts have been made to secure or authenticate access to the simulated quantum processors.
A SimulaQron backend server will by default happily accept requests from any client connecting to it, and in particular allow access to any qubit given its correct identifiers.
It is clear that mechanisms for implementing access control to quantum nodes and qubits are important to realize in software for a quantum internet, highlighting the need for a tool like SimulaQron as a stand-in for quantum hardware in order to develop them.

\subsection{Related Work}
There is to our knowledge no analogue of SimulaQron available for application development on a quantum internet.
There are of course numerous simulators of a quantum computers 
available, some of which could in principle be used as a local simulation backend in conjunction with SimulaQron, with the distributed simulation handled by SimulaQron.
Freely available ones include ProjectQ~\cite{Steiger2016} which is written in Python and contains an optimizing compiler.
Other simulators include Liquid~\cite{Wecker2014} (available as a binary), Forest by Rigetti~\cite{rigetti}, QX~\cite{Khammassi2017}, the simulation backend of the IBM Quantum Experience~\cite{qexperience} (can be accessed in Python using QISKit), and the Microsoft Quantum Development Kit~\cite{microsoftQDK} (using the programming language Q\#).
Presently, the quantum backend of SimulaQron is simply realized using QuTip~\cite{Johansson2012}, which 
is not designed to be an efficient simulator, but more than sufficient for the purpose of simulating a small network.
Using QuTip also has the advantage of providing a very easy way to extend SimulaQron to operate on noisy instead of perfect qubits, should one wish to gain 
insights into whether, for example, software performing error correction for a quantum protocol has been implemented correctly. Again, we emphasize that while this allows verification of the correct implementation of such error correction schemes, SimulaQron is not meant to accurately model time dependent noise in the network.

\subsection{Overview}
\begin{figure}[h]
    \centering
    \includegraphics[width=0.4\textwidth]{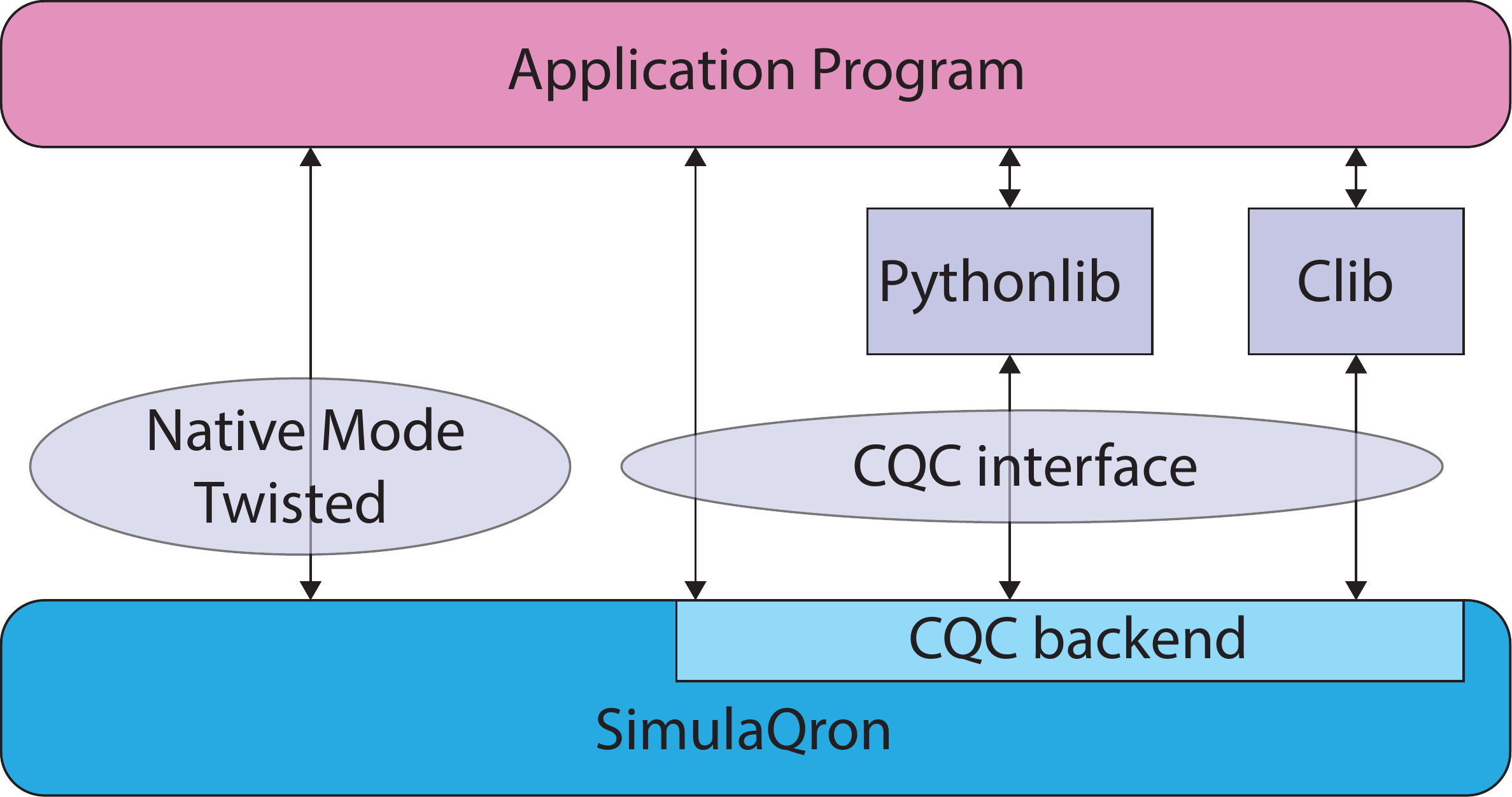}
    \caption{SimulaQron's simulated quantum internet can be programmed in two ways.
        The first proceeds by directly accessing the SimulaQron backend using Twisted in Python.
        This way of programming
        gives full access to the backend, but is highly specific to Twisted and unlikely to be available on any real quantum network devices of the future.
        The second way of programming SimulaQron is by using the Classical-Quantum-Combiner (CQC) interface.
        CQC specifies a packet format for issuing commands to 
        a quantum network node and we intend to make a (possibly refined) version of CQC available on the planned 2020 quantum internet demonstrator in the Netherlands using actual quantum network hardware.
        For ease of programming, we provide two libraries that encapsulate the CQC interface.
        More specifically, we provide libraries in both Python and C for programming quantum internet applications, which internally connect to SimulaQron over the CQC interface.
        Libraries for other languages are easy to add and can access SimulaQron using the CQC interface.}
    \label{fig:cqc_schematic}
\end{figure}

SimulaQron itself is written in Python~\cite{python} due to the popularity of this language in the quantum information community, making it easier to extend if desired.   
Internally, SimulaQron also makes use of the Twisted framework for Python~\cite{twisted}. Twisted is a library providing functions that facilitate the
development of network applications in Python.
SimulaQron's simulated quantum internet can be programmed in two ways as outlined in Figure~\ref{fig:cqc_schematic}.
In Section~\ref{sec:backend} we will provide an overview of the design and inner workings of SimulaQron.
In Section~\ref{sec:cqc} we discuss the integration with the CQC interface.
We provide two simple examples on how to program SimulaQron using the Python CQC library in Section~\ref{sec:example}, together with performance analysis of SimulaQron for some test-cases.
Further examples, programming templates as well as an API documentation can be found online~\cite{simulaqron}. 
Finally, we discuss future developments and extensions of SimulaQron.

\section{Backend}\label{sec:backend}

Let us now first describe the SimulaQron backend, which can be accessed in Python using Twisted's Perspective Broker~\cite{twisted}. 
We note that for programming applications there is no need to ever access the backend directly, as the Python library using CQC provides a much easier
way to program applications. CQC is an interface (close to) what we intend to make available on the quantum internet demonstrator in the Netherlands that can be accessed from any programming language.

\subsection{Challenges}
The main challenge in providing a simulator suitable for programming quantum internet applications is to simulate quantum entanglement. Mathematically, any quantum state can be written as a density matrix $\rho \in \Complex^{d\times d}$, where $d$ is the dimension of the quantum system (see~\cite{Nielsen2010}, or~\cite[Weeks 0 and 1]{mooc} for an introduction to quantum information). Crucially, for two entangled qubits $A$ and $B$, we have $\rho \neq \sum_j p_j \rho^A_j \otimes \rho^B_j$, with
$\rho^A_j \in \Complex^{2 \times 2}$, $\rho^B_j \in \Complex^{2 \times 2}$, and $\otimes$ denoting the tensor product. This means that we cannot factorize $\rho$ into different components $\rho^A_j$ and $\rho^B_j$ that could be simulated individually on two different network nodes. Instead, we need to simulate the entire matrix $\rho$ as one, while making qubits $A$ and $B$ virtually available at two different nodes in the network. 

One way of achieving this, is to let one simulating node hold all qubits in the network in the same register, consisting of a matrix of dimension $2^n \times 2^n$ where $n$ is the number of qubits. This places a prohibitively large load on the simulating node, preventing large networks to be simulated. 
One step towards making the simulation more efficient is to still use only one central simulating node, but keep - as much as possible - different registers. For example, by initially placing each qubit in its own register requiring the simulation of only a $2\times 2$ matrix per initial qubit, and only merging registers as qubits become entangled with each other. This improvement is indeed implemented in SimulaQron whenever qubits are created at one node. Nevertheless if this was the only optimization, a significant load 
would remain at one node in the network. 

Consequently, we here go one step further and choose a distributed approach to simulation, described in detail below. In summary, each node keeps a quantum register of its own, in which a number of qubits are simulated. Qubits which are virtually available at other nodes are mapped back to registers simulated at in principle any other node, typically holding the other qubit of any entanglement. This allows the network to grow dynamically, and distributes simulation efforts amongst many computers. It is especially
suited to situations where large amounts of fragmented entanglement exists in the network. This is naturally the case if the simulated network is very large but different subsets of nodes are executing protocols between them at any one time. This is well motivated from the usage of the classical internet, where many subsets of a few nodes each communicate with each other, but we do not see all nodes on the internet engaging in a joint application protocol. Another natural example, which we will
also consider in a performance analysis below, is given by a situation in which a large number of nodes in a ring keeps entanglement with its neighbours - for example to transmit a qubit by forward teleportation along a chain of nodes. Our analysis shows no significant problems in simulating 60 nodes using 120 qubits on a single desktop machine, while if we had placed all 120 qubits in a single register we would need to keep track of a $2^{120}$ dimensional matrix.

While a distributed simulation is more efficient and allows a dynamic growth of the network, it also brings new challenges. Quantum gates performed on two qubits can cause two qubits that were previously unentangled to become entangled. Consequently, if the two qubits were previously simulated in two different registers in the network, then a register merge is required. This is made challenging by the fact, that other nodes may simultaneously try and manipulate qubits simulated in said registers, or even perform operations that require a different register merge. It is in general challenging to manage access to distributed data in the presence of many concurrent requests to use it. SimulaQron solves such concurrency issues 
using a relatively simple but carefully crafted internal locking mechanism. 

\subsection{Design overview}
\begin{figure}[h]
    \centering
    \includegraphics[width=0.47\textwidth]{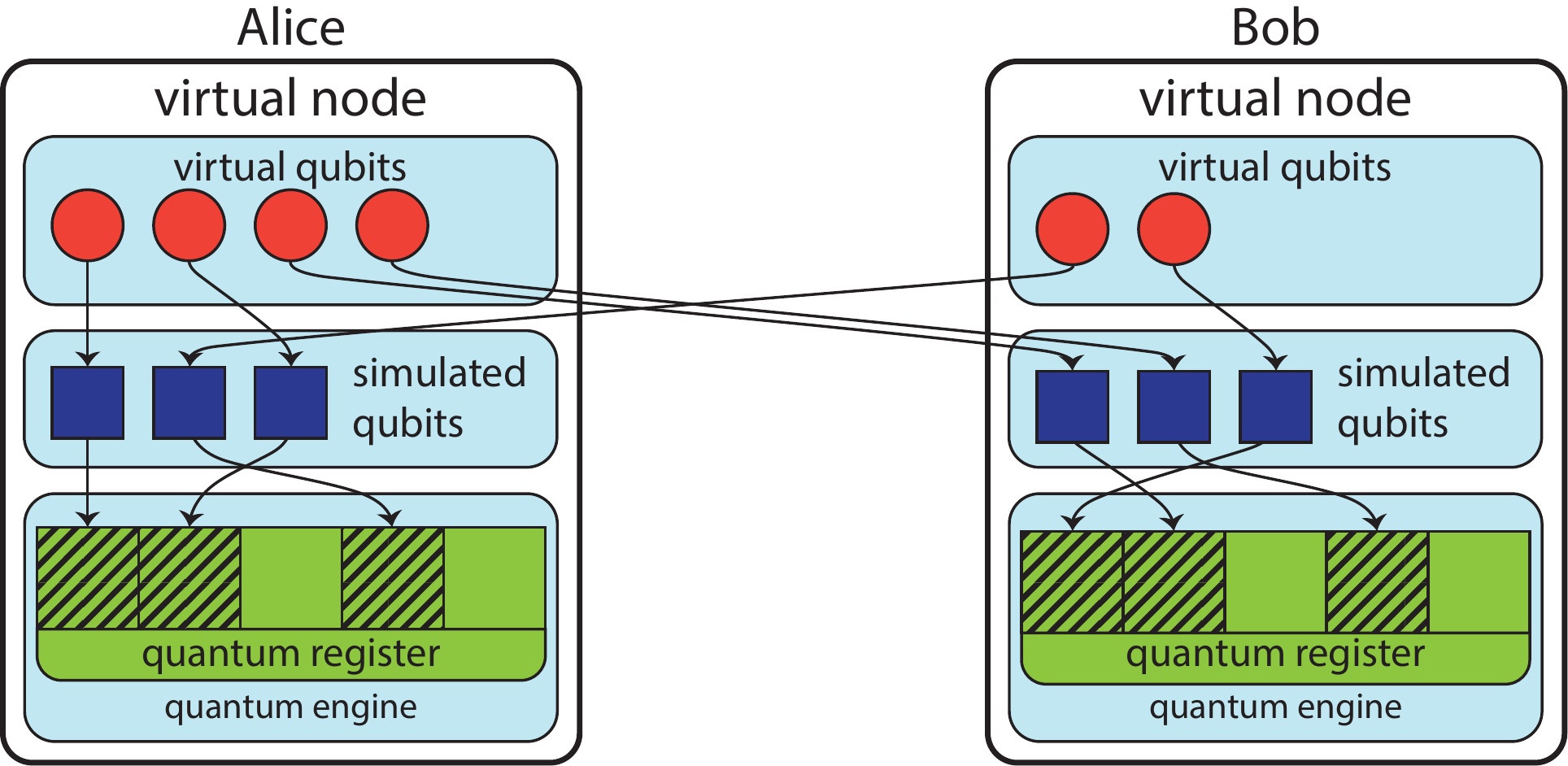}
    \caption{A visualization of the interplay between different internal components of SimulaQron.
        The \inlinecode{simulatedQubit}s (blue squares) are objects handled locally in a \inlinecode{virtualNode}.
        These \inlinecode{simulatedQubit}s point to a part of the \inlinecode{quantumRegister}, which stores the quantum state simulated by the \inlinecode{virtualNode}.
        Operations on the simulated state in the \inlinecode{quantumRegister} are handled by the \inlinecode{quantumEngine}.
        Additionally, a \inlinecode{virtualNode} also has \inlinecode{virtualQubit}s (red circles).
        These \inlinecode{virtualQubit}s point to \inlinecode{simulatedQubit}s, possibly in a different \inlinecode{virtualNode}.
        The \inlinecode{virtualQubit}s correspond to the actual qubits a node would have in a physical implementation of the quantum network.
    }
    \label{fig:simvirtqubits2}
\end{figure}

Together, these considerations motivate the design of the SimulaQron backend depicted in Figure~\ref{fig:simvirtqubits2}. The SimulaQron backend consists of running a client and server program on each classical computer - a \inlinecode{virtualNode} - that wants to participate in the simulated quantum network. All such programs connect to each other, forming the simulated quantum internet backend. 

SimulaQron itself does not provide a new quantum simulator, but rather builds a distributed simulation on top of an existing one using a modular design. In the initial release, we have simply used QuTip for the underlying simulation, but essentially any existing simulator (or even real quantum computing - but not yet networked - hardware) that can be addressed via Python can easily be used in conjunction with SimulaQron. An existing simulator can be made available to SimulaQron as a backend by providing an interface in the form of a \inlinecode{quantumEngine} to function as a \inlinecode{quantumRegister}.
This register supports addressing individual qubits by their position in the register, that is, for a register with $n$ qubits. For example, 
if a Hadamard gate $H$ is requested on qubit $j$, \inlinecode{quantumRegister} is responsible for applying this operation on the underlying
quantum simulation. In the simple case of QuTip, this means simply applying the unitary $\id^{\otimes j-1} \otimes H \otimes \id^{\otimes n-j}$ to the
matrix representing $\rho$ (see \inlinecode{crudeSimulator.py}). 

Building on top of an underlying \inlinecode{quantumRegister}, SimulaQron uses a \inlinecode{simulatedQubit} object to represent each qubit simulated in the underlying \inlinecode{quantumRegister}. Manipulation of qubits then follows exclusively by manipulating such \inlinecode{simulatedQubit}s without interacting directly with the \inlinecode{quantumRegister}. In particular, 
each \inlinecode{simulatedQubit} keeps track of its own position in the register, allowing easy
manipulation and update of qubits that are physically simulated at a particular node. For example, if a qubit is measured it can 
be removed from the underlying \inlinecode{quantumRegister}, effectively shrinking the size of the matrices. This allows the simulation to proceed without having the underlying register grow arbitrarily while new qubits are being created and discarded. Removal from the \inlinecode{quantumRegister} can be done by updating only the simulated qubits (and not the corresponding virtual ones that may be held by other nodes in the simulation), and hence the rest of the simulation can proceed to access the simulated qubits without being aware that the underlying register has been shrunk. Figure~\ref{fig:simvirtqubits2} illustrates the relationship between the \inlinecode{quantumRegister} and the \inlinecode{simulatedQubit}s. We note that \inlinecode{simulatedQubit} objects are local to the node performing their actual simulation in the \inlinecode{quantumRegister}. 

\subsection{Virtual Qubits}

Each \inlinecode{simulatedQubit} object is then associated with a \inlinecode{virtualQubit} object. Importantly, a \inlinecode{virtualQubit} does not need to reside at the same network node as the corresponding \inlinecode{simulatedQubit}, but can be transferred to other nodes than the one performing the backend simulation. Specifically, each \inlinecode{virtualNode} can hold a number of \inlinecode{virtualQubit}s which can either be simulated locally (i.e., the simulated qubit
and the virtual qubit are located at the same node), or at a remote node. Twisted's Perspective Broker is used to marshall the mapping of virtual qubits back to simulated qubit objects at remote nodes on the simulated network. 

This way it is possible for SimulaQron to simulate entanglement between network nodes: If simulated nodes $A$ and $B$ hold entanglement, then each \inlinecode{virtualNode} has a \inlinecode{virtualQubit} that can be processed at nodes $A$ and $B$ as if they were truly entangled. In the background, however, the two virtual qubits are mapped back to two \inlinecode{simulatedQubit}s within the same \inlinecode{quantumRegister} that may be located at either $A$ or $B$, or even at
some other node.

Any application wishing to use SimulaQron consists of a client program at each computer that connects to the SimulaQron server backend. SimulaQron will make virtual qubits available to any client connecting to the \inlinecode{virtualNode} server, as well as allow the creation of new qubits and registers. Using Twisted's Perspective Broker, the client has full access to the specific \inlinecode{virtualQubit}, allowing it to perform gates, measurements, or send the qubit to other nodes in the simulated quantum network.
 
\subsection{Register merges}
One simple way to deal with virtual qubits is to have them simulated at one single classical computer in the network. Evidently, this puts a large strain on that specific computer, which then has to simulate every single qubit in the network. Here, we have instead chosen a distributed simulation, which is more challenging to realize, but does allow many nodes to take part in the simulation, as long as the global entanglement in each quantum state is not too large. This is typically the case, when investigating how to program applications that may each run on only a subset of the nodes in the network.

\begin{figure*}[ht]
    \centering
    \includegraphics[width=0.9\textwidth]{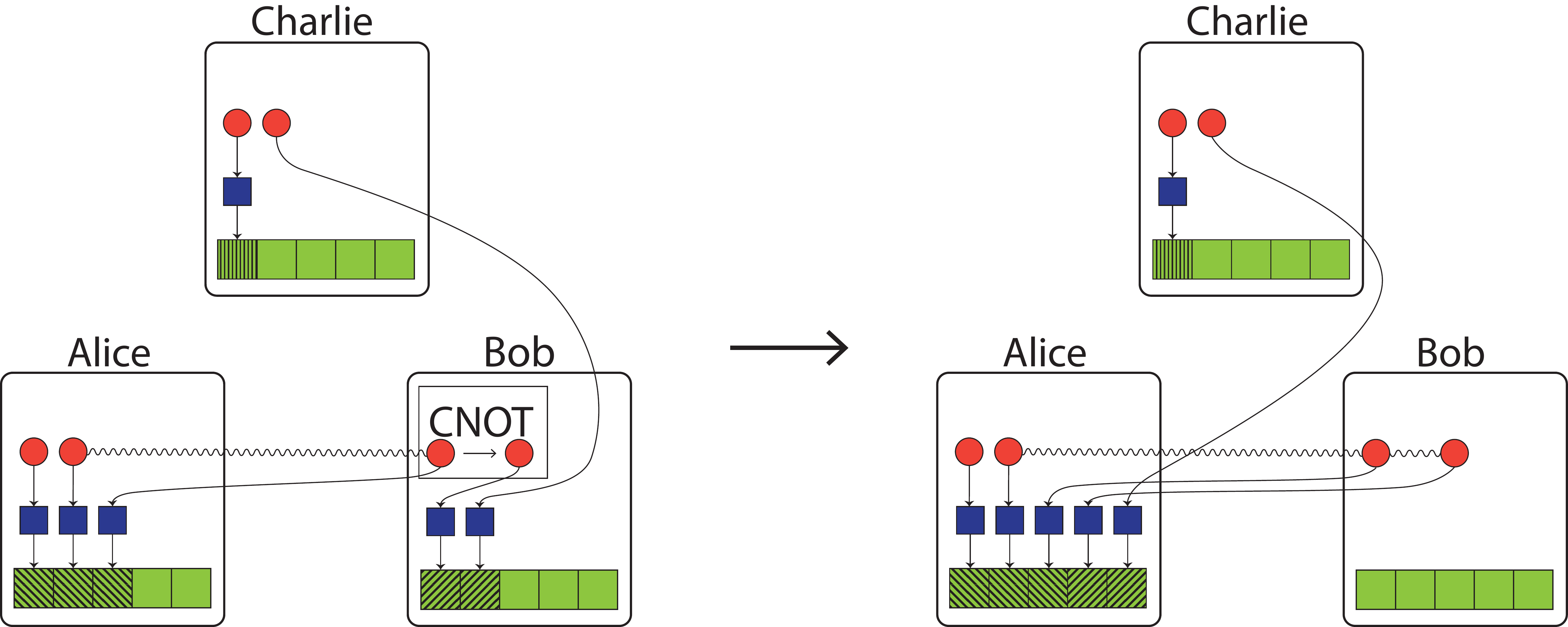}
    \caption{A visualization of a register merge, initialized by a CNOT performed in the \inlinecode{virtualNode} Bob.
        Alice and Bob initially share a EPR-pair, represented by a curvy line between the two \inlinecode{virtualQubit}s.
        These entangled qubits are simulated in the \inlinecode{virtualNode} Alice.
        Bob then performs a CNOT between his part of the entangled state (control) and another qubit (target) simulated at Bob.
        This operations requires a register merge between Alice and Bob, since the three relevant qubits will now be in a GHZ-state, which needs to be simulated locally.
        The registers of Alice and Bob are therefore merged and all qubits simulated by either Alice or Bob are now simulated by Alice.
        Note that this also includes the \inlinecode{virtualQubit} of Charlie, which is initially simulated by Bob.
    }
    \label{fig:register_merge}
\end{figure*}

We hence allow any \inlinecode{virtualNode} to hold a \inlinecode{quantumRegister} containing simulated qubits that may eventually be sent as virtual qubits to other nodes in the network. This enables a distributed simulation in which each classical computer participating in the simulated quantum internet contributes its share to realizing the overall simulation. Yet, it is clear that this approach brings additional challenges. Specifically, consider two virtual qubits $A$ and $B$ at one network node, which are mapped back to two simulated qubits in \emph{different}
\inlinecode{quantumRegister}s. An example of such a situation is given in Figure~\ref{fig:register_merge}. 
If the protocol now requests the \inlinecode{virtualNode} to perform an entangling gate between $A$ and $B$, qubits $A$ and $B$ may now become entangled and can thus no longer be simulated in different registers. 

Entangling gates may thus require a register merge in order for the simulation to proceed. More precisely, 
entangling $A$ and $B$ then requires that the two distinct quantum registers which hold the corresponding simulated qubits to be merged into one single quantum register, followed by an update of the corresponding \inlinecode{simulatedQubit}s that the \inlinecode{virtualQubit}s are mapped to. 

SimulaQron solves this problem by transparently merging the registers in the background. For the application dealing with only the \inlinecode{virtualQubit}s such a merge is invisible, but the simulated qubits representing the virtual qubits in question are updated.

It is clear that locking is required to ensure consistency in performing such register merges in case multiple requests to merge a register arise in the network at the same time. At present, SimulaQron implements a very simple and relatively inefficient locking mechanism tailored to acquiring the minimal set of locks required to perform and update. As two separate locks for registers and simulated qubits are used, a deadlock can arise which is dealt with using a simple randomized backoff to marshall competition for locks. Future versions of SimulaQron may be enhanced by a more sophisticated locking mechanism.

\section{CQC}\label{sec:cqc}
The CQC instruction format provides a low level language programmable at the level of quantum gates and measurements, tailored specifically to include certain commands and behaviour that is useful in the quantum internet domain.
The CQC interface provides a way for applications to be developed independently of the underlying platform and for these to be executed on any platform which provide the CQC interface.
On top of the backend, SimulaQron realizes a CQC interface (classical-quantum combiner) as implemented by the so-called CQC backend.
SimulaQron can therefore be programmed using any language capable of connecting to the CQC server backend over a TCP connection, and sending packets of the required form, specified by the CQC message format.
In this section we describe the CQC interface and its messages in more detail.


Two libraries are included to program SimulaQron via the CQC interface in Python and C. Programming SimulaQron via this Python library is the easiest way to make use of SimulaQron and the recommended way to get started using SimulaQron. 
Below, we provide two simple examples; many more examples can be found in the online documentation~\cite{simulaqron}. 

We expect that the present CQC interface will undergo further evolution until its use in the Dutch demonstration network. As such, we describe the main ideas and what functionality exists. 
The current version of the CQC interface including the precise message format is available online~\cite{simulaqron} and will be updated when a newer version exist.
CQC specifies a particular instruction format for requests and replies from the CQC backend. Requests may be given as a single message to the CQC interface, or an entire batch of messages at once specifying a complex operation consisting of many actions in succession. 

Messages in the CQC interface can be of different \emph{types}, as described in section \ref{sec:CQC_types}. 
One of these types is \inlinecode{COMMAND} which instructs SimulaQron (or the physical hardware) to perform a certain command. This command could for example be a gate or to send a qubit to another node.
More details on the message type \inlinecode{COMMAND} is given in section \ref{sec:CQC_CMD}.

In addition to the type of message and what command to perform, CQC messages will also contain information regarding for example the CQC version, an application identifier, the length of the message (including additional commands if any) and - if applicable - the qubit identifiers a command concerns, the IP and port-number the node a qubit should be sent to.
Other options that can be specified is whether notification should be returned when the command is finished and if the node should be blocked during the execution of the command, see table \ref{tab:options}.

\begin{table}[h]
    \begin{tabular}{|l|l|}
        \hline
        \textbf{Option}            & \textbf{Effect}                        \\ \hline
        \inlinecode{OPT_NOTIFY}    & Send a notify when command is done            \\ \hline
        \inlinecode{OPT_ACTION}    & Execute further commands when done \\ \hline
        \inlinecode{OPT_IFTHEN}    & Execute further commands based on result \\ \hline
        \inlinecode{OPT_BLOCK}     & Block until command done                   \\ \hline
    \end{tabular}
    \caption{Options that can be specified when sending a \inlinecode{COMMAND} message.}
    \label{tab:options}
\end{table}

There is also the possibility to attach a list of commands that should be executed directly after the command specified in the CQC message is completed, or if, for example, a measurement outcome takes on a certain value. As such, it is assumed that the CQC backend provides a rudimentary form of classical logic next to the quantum specific instructions that allow certain simple processing to be executed in the CQC backend, or indeed the actual hardware. 
That is, these commands can be executed without having to send messages back and fourth through the CQC interface between each command.
Avoiding these messages between each command allows the sequence of commands to be executed faster, which is important for a physically implemented quantum network, subject to decoherence.

\subsection{Message types}\label{sec:CQC_types}
We describe in this section message types currently implemented in the CQC interface.
As mentioned above, in future version of the interface, changes may be made and the latest version can be found online.
The types of messages sent from the application to the hardware (either simulated by SimulaQron or physically implemented) are specified in table \ref{tab:types_down}.

\begin{table}[h]
    \begin{tabular}{|l|l|}
        \hline
        \textbf{Type}            & \textbf{Effect}                          \\ \hline
        \inlinecode{TP_HELLO}       & Alive check (get hardware spec.)      \\ \hline
        \inlinecode{TP_COMMAND}     & Execute (list of) command             \\ \hline
        \inlinecode{TP_FACTORY}     & Execute (list of) command repeatedly  \\ \hline
        \inlinecode{TP_GET_TIME}    & Get creation time of qubit            \\ \hline
    \end{tabular}
    \caption{Types of messages from the application to the hardware/SimulaQron.}
    \label{tab:types_down}
\end{table}

A \inlinecode{HELLO} message can be used to check that the connection to the hardware/SimulaQron is up and also to get some specifications about the hardware.
The commands that can be specified using the type \inlinecode{COMMAND} is discussed in the following section.
\inlinecode{FACTORY} is a type similar to \inlinecode{COMMAND}, but here a certain command should be executed a specified number of times.
An example of the use of \inlinecode{FACTORY} is to instruct the hardware to continuously produce a specified number of entangled EPR-pairs.
An EPR-pair could then ready to be used whenever it is needed by the protocol running on the network.

There are also message types returned by the hardware to the application, as specified in table \ref{tab:types_up}.

\begin{table}[h]
    \begin{tabular}{|l|l|}
        \hline
        \textbf{Type}                 & \textbf{Effect}                  \\ \hline
        \inlinecode{TP_NEW_OK}        & Qubit was allocated              \\ \hline
        \inlinecode{TP_EXPIRE}        & Qubit has expired                \\ \hline
        \inlinecode{TP_DONE}          & Done with command                \\ \hline
        \inlinecode{TP_RECV}          & Received qubit                   \\ \hline
        \inlinecode{TP_EPR_OK}        & Created EPR pair                 \\ \hline
        \inlinecode{TP_MEASOUT}       & Measurement outcome              \\ \hline
        \inlinecode{TP_INF_TIME}      & Return timing information        \\ \hline
        \inlinecode{ERR_GENERAL}      & General purpose error            \\ \hline
        \inlinecode{ERR_NOQUBIT}      & No more qubit                    \\ \hline
        \inlinecode{ERR_UNKNOWN}      & Unknown qubit ID                 \\ \hline
        \inlinecode{ERR_UNAVAILABLE}  & Cannot allocate qubit            \\ \hline
        \inlinecode{ERR_DENIED}       & No access to qubit               \\ \hline
        \inlinecode{ERR_VERSION}      & CQC version not supported        \\ \hline
        \inlinecode{ERR_UNSUPP}       & Sequence not supported           \\ \hline
        \inlinecode{ERR_TIMEOUT}      & Timeout                          \\ \hline
    \end{tabular}
    \caption{Types of messages from the hardware/SimulaQron to the application.}
    \label{tab:types_up}
\end{table}

\subsection{Possible commands}\label{sec:CQC_CMD}
The different commands currently supported in the CQC interface when using the message type \inlinecode{COMMAND} are specified in table \ref{tab:cmd}.
The full CQC packet format includes qubit and entanglement identifiers required by the commands below~\cite{simulaqron}. We remark that we include a command to create pairwise entanglement, since this reflects the way two nodes are entangled by a heralded entanglement generation scheme (see Figure~\ref{fig:cqc_simple} for a high level implementation schematic and information).
When a qubit is received or entanglement has been generated, a message, \inlinecode{TP_RECV} or \inlinecode{TP_EPR_OK} respectively, is returned to the application.
This message notifies the application that the command was successful and that possible corrections required by the entanglement generation scheme has been applied.

\begin{table}[h]
    \begin{tabular}{|l|l|}
        \hline
        \textbf{Command}                  & \textbf{Effect}                                     \\ \hline
        \inlinecode{CMD_NEW}              & Ask to allocate a new qubit                         \\ \hline
        \inlinecode{CMD_ALLOCATE}         & Ask to allocate multiple qubits                     \\ \hline
        \inlinecode{CMD_RELEASE}          & Release qubit to be used by other app.              \\ \hline
        \inlinecode{CMD_RESET}            & Reset qubit to $\ket{0}$                            \\ \hline
        \inlinecode{CMD_MEASURE}          & Measure qubit (demolition)                          \\ \hline
        \inlinecode{CMD_MEASURE_INPLACE}  & Measure qubit (non-demolition)                      \\ \hline
        \inlinecode{CMD_SEND}             & Send qubit to another node                          \\ \hline
        \inlinecode{CMD_RECV}             & Ask to receive qubit                                \\ \hline
        \inlinecode{CMD_EPR}              & Create EPR pair with another node                   \\ \hline
        \inlinecode{CMD_RECV_EPR}         & Ask to receive half of EPR pair                     \\ \hline
        \inlinecode{CMD_SWAP} 		      & Entanglement swapping                               \\ \hline
        \inlinecode{CMD_I}                & Identity                                            \\ \hline
        \inlinecode{CMD_X}                & Pauli X                                             \\ \hline
        \inlinecode{CMD_Y}                & Pauli Y                                             \\ \hline
        \inlinecode{CMD_Z}                & Pauli Z                                             \\ \hline
        \inlinecode{CMD_H}                & Hadamard                                            \\ \hline
        \inlinecode{CMD_K}                & K gate (Z to Y)                                     \\ \hline
        \inlinecode{CMD_T}                & T gate                                              \\ \hline
        \inlinecode{CMD_ROT_X}            & Rotation around X (multiple of $\frac{2\pi}{256}$)  \\ \hline
        \inlinecode{CMD_ROT_Y}            & Rotation around Y (multiple of $\frac{2\pi}{256}$)  \\ \hline
        \inlinecode{CMD_ROT_Z}            & Rotation around Z (multiple of $\frac{2\pi}{256}$)  \\ \hline
        \inlinecode{CMD_CNOT}             & CNOT   (this qubit as control)                      \\ \hline
        \inlinecode{CMD_CPHASE}           & CPHASE (this qubit as control)                      \\ \hline
    \end{tabular}
    \caption{Commands that can be specified when using the type \inlinecode{COMMAND}.}
    \label{tab:cmd}
\end{table}

The angle of rotation for the single-qubit rotations are currently specified by one byte and is therefore discretized to 256 possible angles.
When the creation of an EPR pair is requested, a message containing an entanglement identifier will be returned to the application. 
At present, there is no protocol to track and manage entanglement in a network available. We are currently developing and testing such a protocol, which will be referenced on the SimulaQron website once released~\cite{simulaqron}.

\section{Examples}\label{sec:example}
We provide here a simple example of how to program SimulaQron using the Python CQC library (see Figure~\ref{fig:cqc_schematic}).
More examples, also using the C CQC library, can be found in the full online documentation~\cite{simulaqron}.
Before running the example presented here, the simulated network we consider needs to be configured and the severs that does the communication in the backend of SimulaQron has to be setup.
Information on how to configure the simulated network and setup the servers can be found in the online documentation~\cite{simulaqron}.
In what follows, we will hence assume that the SimulaQron and CQC backends have been setup already, simulating the hardware of two quantum internet nodes labelled Alice and Bob. We remark that the names Alice and Bob are translated by the Python CQC library into IP addresses according the configuration file specified on~\cite{simulaqron}.

\subsection{Sending BB84 States}

The first example we consider has no classical communication on the application level between Alice and Bob. This implies that we do not need to set up a
separate client/server interaction to exchange information at the level of applications. This means that Alice and Bob only connect locally to the CQC backend of SimulaQron, which functions as the quantum hardware at their network node, to issue quantum instructions. 
In this example, Alice will send a single qubit to Bob. 

\subsubsection{Code for Alice}
The first thing we need to do is to initialize an object called a \inlinecode{CQCConnection}, which does the communication with SimulaQron using the CQC interface.
Once this connection is set up, Alice has access to her own locally simulated quantum hardware. 
\begin{lstlisting}[language=Python,basicstyle=\footnotesize\ttfamily]
    from SimulaQron.cqc.pythonLib.cqc import *

    # Establish connection to SimulaQron
    Alice=CQCConnection("Alice")
\end{lstlisting}
The argument that \inlinecode{CQCConnection} takes should be the name specified in the configuration file for the CQC-network.
Once the connection is set up we can then create our first qubit.
The \inlinecode{qubit}-object takes a \inlinecode{CQCConnection} as argument when initialized.
When operations are applied to the \inlinecode{qubit}, the \inlinecode{CQCConnection} is used to communicate with SimulaQron, such that the corresponding simulatedQubit is updated.
\begin{lstlisting}[language=Python,basicstyle=\footnotesize\ttfamily]

    # Create new qubit
    q=qubit(Alice)
\end{lstlisting}
Alice will then send Bob one out of the four states
\begin{equation}\label{eq:four_states}
    \ket{0},\quad\ket{1},\quad\ket{+},\quad\ket{-}
\end{equation}
where $\ket{\pm}=\frac{1}{\sqrt{2}}(\ket{0}\pm\ket{1})$, depending on the choice of the bits $h_A$ and $x$ set in the program.
The states in equation~\eqref{eq:four_states} are usually called BB84-states and can be described by two binary variables as follows
\begin{equation}
    H^{h_A}X^{x}\ket{0}\qquad h_A,x\in\{0,1\},
\end{equation}
where $H$ is the Hadamard operation and $X$ is the Pauli-X operation, defined as
\begin{equation}
    H=\frac{1}{\sqrt{2}}\begin{pmatrix} 1 & 1 \\ 1 & -1 \end{pmatrix},\qquad X=\begin{pmatrix} 0 & 1 \\ 1 & 0 \end{pmatrix}.
\end{equation}
The second part of the code on Alice's side will then apply the operation $H^{h_A}X^{x}$ and send the qubit to Bob. In our simple example, Alice is done
using the quantum internet and hence we close the \inlinecode{CQCConnection}.
\begin{lstlisting}[language=Python,basicstyle=\footnotesize\ttfamily]

    # Determine which BB84 state to use
    h_A=1;x=0

    # if x=1, flip |0> to |1>
    if x == 1: q.X()

    # if h_A==1, convert to Hadamard basis
    if h_A==1: q.H()

    # Send qubit to Bob
    Alice.sendQubit(q,"Bob")

    # Close connection to SimulaQron
    Alice.close()
\end{lstlisting}

\subsubsection{Code for Bob}
We will now describe the code on Bobs side. In our example, Bob does nothing but wait for a qubit to arrive. Once he receives one, 
he will measure it in the standard- ($\{\ket{0},\ket{1}\}$) (for $h_B = 0$) 
or the Hadamard-basis ($\{\ket{+},\ket{-}\}$) (for $h_B=1$) and print the measurement outcome.
The code on Bobs side can be seen below, where \inlinecode{h_B} determines that basis Bob measures in.
\begin{lstlisting}[language=Python,basicstyle=\footnotesize\ttfamily]
    from SimulaQron.cqc.pythonLib.cqc import *

    # Establish a connection to SimulaQron
    Bob=CQCConnection("Bob")

    # Choose which basis we measure in
    h_B=1

    # Wait to receive a qubit
    q=Bob.recvQubit()

    # If we chose the Hadamard basis 
    # to measure in, apply H
    if h_B==1: q.H()

    # Measure the qubit in the standard 
    # basis and store the outcome in m
    m=q.measure()

    # Print measurement outcome
    print("Bobs meas. outcome: {}".format(m))

    # Close connection to SimulaQron
    Bob.close()
\end{lstlisting}
Note that if $h_A=h_B$ the measurement outcome for Bob will be Alice's choice of $x$ with probability one.

\subsection{Teleporting a qubit}
In our second example, Alice teleports a qubit to Bob.
This example demonstrates how to create shared EPR pairs, and also how to perform additional classical communication from Alice and Bob on the application level. This classical communication can be realized using standard client/servers programming. An inefficient, but convenient testing tool, to perform classical communication tool is provided by the Python library that does not
require knowledge of classical client/server programming.

\subsubsection{Code for Alice}
As in the previous example, the first thing that happens is that a \inlinecode{CQCConnection} is initialized to handle the communication to SimulaQron via the CQC interface.
\begin{lstlisting}[language=Python,basicstyle=\footnotesize\ttfamily]
    # Initialize the connection
    Alice=CQCConnection("Alice")
\end{lstlisting}
By calling the method \inlinecode{createEPR}, Alice makes a request to generate an EPR-pair with Bob, i.e. to generate the state $\frac{1}{\sqrt{2}}\left(\ket{00}_{AB}+\ket{11}_{AB}\right)$.
\begin{lstlisting}[language=Python,basicstyle=\footnotesize\ttfamily]
    # Make an EPR pair with Bob
    qA=Alice.createEPR("Bob")
\end{lstlisting}
When the EPR-pair has been generated, Alice prepares another qubit \inlinecode{q} in the state $\ket{+}=H\ket{0}$, which she wants to teleport to Bob.
\begin{lstlisting}[language=Python,basicstyle=\footnotesize\ttfamily]
    # Create a qubit to teleport
    q=qubit(Alice)

    # Prepare the qubit to teleport in |+>
    q.H()
\end{lstlisting}
Alice makes a Bell-measurement between the qubit \inlinecode{qA} in the EPR-pair shared with Bob and the qubit \inlinecode{q}.
The Bell-measurement is done by applying a CNOT gate with \inlinecode{q} as control and \inlinecode{qA} as target, followed by a Hadamard gate on \inlinecode{q} and measuring both qubits in the standard basis.
\begin{lstlisting}[language=Python,basicstyle=\footnotesize\ttfamily]
    # Apply the local teleportation operations
    q.cnot(qA)
    q.H()

    # Measure the qubits
    a=q.measure()
    b=qA.measure()
    print("Alice meas. out.: a={}, b={}".format(a,b))
\end{lstlisting}

At this point Alice needs to communicate to Bob what her measurement outcomes was, such that Bob can recover the state which is teleported.
This classical communication can be realized by setting up a client/server interaction between Alice and Bob.
There is a built-in feature in the Python library that realize this functionality, which have been developed for ease of use for someone not familiar with a client/server setup.
This communication is also handled by the object \inlinecode{CQCConnection}.
For Alice to sent a message to Bob, the method \inlinecode{Alice.sendClassical("Bob",msg)} is simple called, where \inlinecode{msg} is the message she wish to send to Bob.
The method opens a socket connection to Bob, sends the message and closes the connection again.
Note that if this method is never called, a socket connection is never opened.

We emphasise that to have classical communication between the applications, one is not forced to use the built-in functionality realized by the \inlinecode{CQCConnection}.
A standard client/server setup can also be used.

\begin{lstlisting}[language=Python,basicstyle=\footnotesize\ttfamily]
    # Send corrections to Bob
    Alice.sendClassical("Bob",[a,b])

    # Stop the connections
    Alice.close()
\end{lstlisting}

\subsubsection{Code for Bob}
As mentioned, Bob will need to know the measurement outcomes from Alice and will therefore setup a server to be able to receive these.
Bob will then receive the qubit \inlinecode{qB} which is part of the EPR-pair generated with Alice.
By calling the method \inlinecode{recvClassical}, Bob receives the measurement outcomes that Alice sent.
Corrections are then performed, depending on these measurement outcomes.
The qubit \inlinecode{qB} will then be in the state Alice prepared, i.e. the state $\ket{+}$.
Finally, Bob measures the qubit \inlinecode{qB} which gives $0$ or $1$ with equal probability.

\begin{lstlisting}[language=Python,basicstyle=\footnotesize\ttfamily]
    # Initialize the connection
    Bob=CQCConnection("Bob")

    # Make an EPR pair with Alice
    qB=Bob.recvEPR()

    # Receive info about corrections
    data=Bob.recvClassical()
    message=list(data)
    a=message[0]
    b=message[1]

    # Apply corrections
    if b==1: qB.X()
    if a==1: qB.Z()

    # Measure qubit
    m=qB.measure()
    print("Bob meas. out.: {}".format(m))

    # Stop the connection
    Bob.close()
\end{lstlisting}

\subsection{Performance}\label{sec:performance}
\begin{figure*}[ht]
    \centering
    \includegraphics[width=0.9\textwidth]{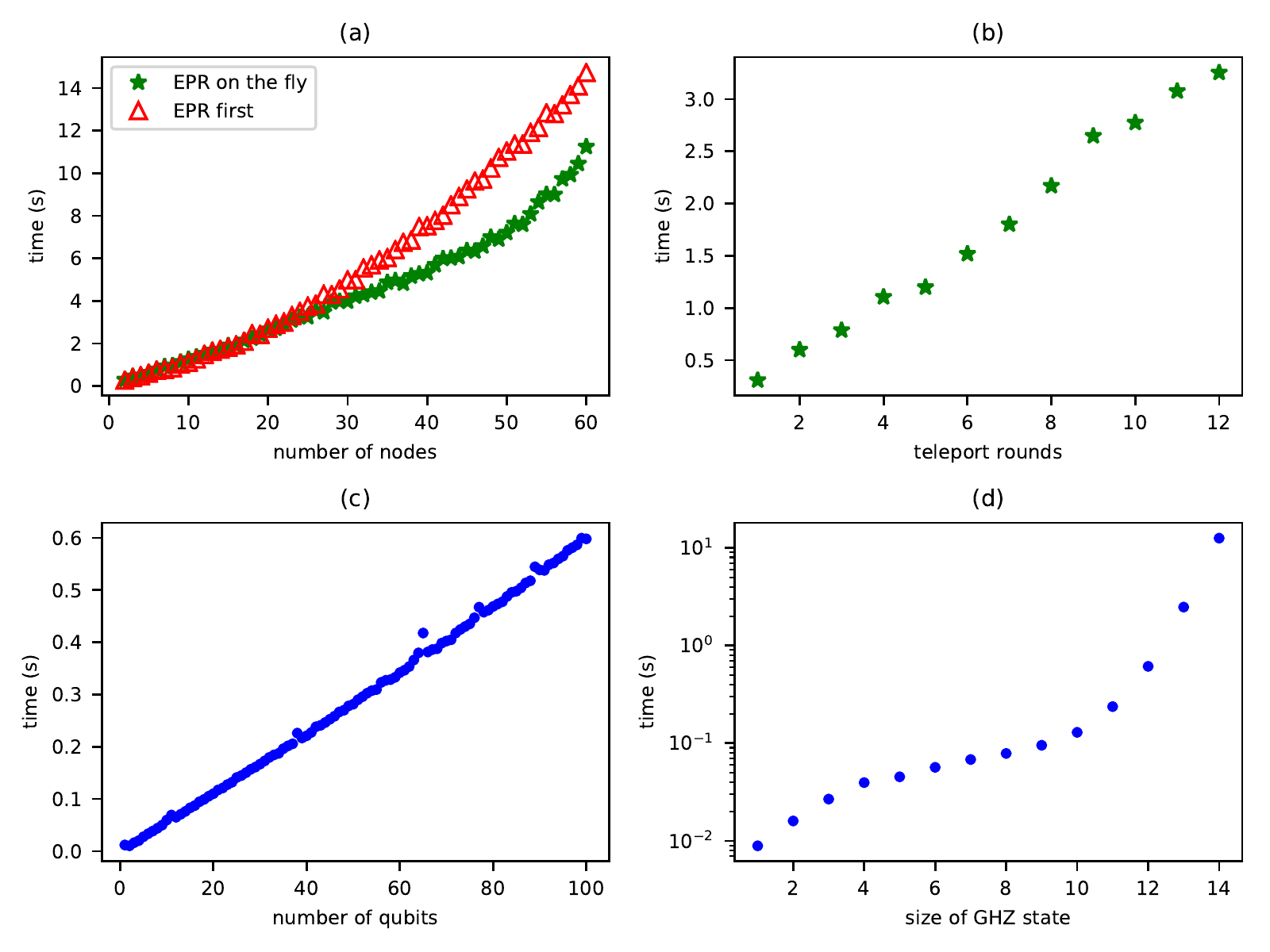}
    \caption{
        Runtimes for the four tests (see Section~\ref{sec:performance}).
        All the tests were executed on one (two for (b)) iMac with a 3.2 GHz Intel Core i5 processor with 8 GB of 1600 MHz RAM running Python 3.6.0 with Twisted 17.9.0.
    }
    \label{fig:performance}
\end{figure*}

In this section we present a practical performance analysis of SimulaQron, which has been performed by running the following four tests:
\begin{enumerate}[label=(\alph*)]
    \item A network with $n$ nodes in a ring is used to teleport a qubit $n$ times from a sender, through all the nodes and back to the sender again.
        The sender records the time it took for the qubit to traverse the network.
        We test two different cases, depending on when the nodes create the EPR-pairs for teleportation: 
(1) Each node creates an EPR-pair with the next node on the ring only once the qubit to be teleported is received.
(2) The creation of all EPR-pairs starts in advance. The qubit is teleported forward as soon as the next EPR-pair becomes available.
        Case (1) is denoted "EPR on the fly" in Figure~\ref{fig:performance} and case (2) as "EPR first".
	This test is executed on the same computer.
    \item A network consisting of two nodes is used to teleport a qubit back and forth between the nodes $n$ times.
        This means teleporting a qubit $2n$ times.
        In this test each node is simulated on its own physical computer, which both are on the same Ethernet network.
        The time it takes to perform all the teleportations of the qubit is again recorded.
    \item Here we test how much time it takes for one node to initialize $n$ qubits and later measure them.
        No two-qubit gate, that can entangle the qubits, is used in this test.
    \item One node initializes a GHZ state on $n$ qubits, i.e. $\ket{\mathrm{GHZ}_n}=\frac{1}{\sqrt{2}}(\ket{0}^{\otimes n}+\ket{1}^{\otimes n})$, and measures all the qubits.
        The time this takes is recorded.
\end{enumerate}

The runtime for these tests can be found in Figure~\ref{fig:performance}.
Note that for these simulations there are three processes running for each node: one performing the actual simulation, one listening to incoming CQC messages and the application program which sends the CQC messages, see Figure~\ref{fig:servers}.
Thus, in the case of 60 nodes in test (a), there are in fact 180 processes running on a single computer.
These processes communicate over TCP, which creates some delay in the runtime of the simulation.

Since \inlinecode{quantumRegister}s are only merged when needed, i.e when a two-qubit gate is performed between qubits in different \inlinecode{quantumRegister}s, the runtime highly depends on how large multi-partite entangled states are produced in the simulation.
For example, in test (c) non-entangled qubits are generated in one node and the runtime then scales linearly with the number of qubits.
On the other hand in test (d), all qubits are in a single entangled state which requires all qubits to be in the same \inlinecode{quantumRegister} and the runtime then scales super-exponentially with the number of qubits.
We emphasize that this runtime 
is a direct consequence of using QuTip as a backend for the simulation.
Creating a GHZ-state directly in QuTip gives the same runtime-scaling as in Figure~\ref{fig:performance}(d).
Thus, by using a different backend for the simulation, the runtime to create larger entangled states can certainly be improved.

As mentioned, new qubits are always put in different registers and these registers are only merged if a two-qubit gate is performed between qubits in different registers.
Currently fixed decisions are used to decide which node will store the register after the merge, depending on which qubit is the target and which is control in the two-qubit gate.
There is clearly a room for improvement of the performance, if the direction of the register merge is choosen in a more dynamic way, depending on the protocol being simulated.
At this point we do not have enough usage data to know what good decisions for the direction of the register merge are.
We also point out that we make no effort in trying to split registers if qubits later become unentangled by some other gate, since this is unlikely to be 
feasible in general and would require a lot of extra computations.

We emphasize that SimulaQron is not intended to substitute simulators dedicated to handle large multipartite states, as in the case for quantum computation.
Despite this, developing software realizing for example distributed quantum computation using SimulaQron, is in principle possible since universal quantum computation is supported by the provided operations.
However, SimulaQron's use-cases are for larger (or smaller) networks where far from all nodes are entangled in a single state.
Nonetheless, this does not exclude cases where each node share some entanglement with another node, for example as in test (a) where all the nodes share two EPR-pairs with two other nodes, since these EPR-pairs can be simulated in different \inlinecode{quantumRegisters} and does not require a single \inlinecode{quantumRegister} for the whole network.

\section{Conclusion}\label{sec:future}
SimulaQron provides access to simulated quantum internet hardware, enabling application development. SimulaQron is undergoing continuous development and improvements and new features may be added over time. 

CQC itself can be understood as a very low level language by itself, supported directly by SimulaQron, and tailored specifically to the needs and behaviour of quantum internet application programs.  At present, CQC is hidden away from view by the Python and C libraries that enable application development at a much higher level. Indeed, programming using the Python CQC library is the fastest way to get started writing a quantum internet application using SimulaQron. It is conceivable that CQC will be integrated into higher level languages directly by a means of a compiler, which produces CQC commands instead of using a dedicated library.

We might also imagine that more convenient higher level programming environments and libraries become available in the future. While the Python CQC library already offers many conveniences, such as dealing with qubits as Python objects, performing specific gates to - for example - generate BB84 states, it is still a relatively low level approach to programming the quantum internet. Indeed, many protocols depend on exchanging BB84 states, and their creation and processing could in the future be handled by 
a set of quantum internet software libraries.

\acknowledgments

We thank Ronald Hanson, Tracy Northup, Peter Humphreys, Norbert Kalb and Arie van Deursen for interesting discussions and feedback.
AD and SW were supported by STW Netherlands, and NWO VIDI grant, and an ERC Starting grant.

\bibliographystyle{unsrt}
\bibliography{simulaqron,websites}

\begin{thebibliography}{10}

\bibitem{bb84}
Charles~H Bennett and Gilles Brassard.
\newblock {Quantum Cryptography: Public Key Distribution, and Coin-Tossing}.
\newblock In {\em Proc. 1984 IEEE International Conference on Computers,
  Systems, and Signal Processing}, pages 175--179, 1984.

\bibitem{e91}
Artur~K. Ekert.
\newblock {Quantum cryptography based on Bell's theorem}.
\newblock {\em Physical Review Letters}, 67(6):661--663, aug 1991.

\bibitem{secureID}
Ivan Damg{\aa}rd, Serge Fehr, Louis Salvail, and Christian Schaffner.
\newblock {Secure identification and QKD in the bounded-quantum-storage model}.
\newblock {\em Theoretical Computer Science}, 560(P1):12--26, dec 2014.

\bibitem{Kaniewski2016}
Jedrzej Kaniewski and Stephanie Wehner.
\newblock {Device-independent two-party cryptography secure against sequential
  attacks}.
\newblock {\em New Journal of Physics}, 18(5), 2016.

\bibitem{Jozsa2000}
Richard Jozsa, Daniel~S. Abrams, Jonathan~P. Dowling, and Colin~P. Williams.
\newblock {Quantum Clock Synchronization Based on Shared Prior Entanglement}.
\newblock {\em Physical Review Letters}, 85(9):2010--2013, aug 2000.

\bibitem{Childs2001}
Andrew~M Childs.
\newblock {Secure assisted quantum computation}.
\newblock {\em Quantum Information {\&} Computation}, 5(6):456--466, nov 2001.

\bibitem{telescope}
A.~Kellerer.
\newblock {Quantum telescopes}.
\newblock {\em Astronomy and Geophysics}, 55(3):1--12, 2014.

\bibitem{Muralidharan2016}
Sreraman Muralidharan, Linshu Li, Jungsang Kim, Norbert L{\"{u}}tkenhaus,
  Mikhail~D. Lukin, and Liang Jiang.
\newblock {Optimal architectures for long distance quantum communication}.
\newblock {\em Scientific Reports}, 6(February):1--10, 2016.

\bibitem{VanMeter2014}
Rodney {Van Meter}.
\newblock {\em {Quantum Networking}}.
\newblock John Wiley {\&} Sons, Ltd, Chichester, UK, apr 2014.

\bibitem{Sangouard2011}
Nicolas Sangouard, Christoph Simon, Hugues {De Riedmatten}, and Nicolas Gisin.
\newblock {Quantum repeaters based on atomic ensembles and linear optics}.
\newblock {\em Reviews of Modern Physics}, 83(1):33--80, 2011.

\bibitem{Azuma2015}
Koji Azuma, Kiyoshi Tamaki, and Hoi~Kwong Lo.
\newblock {All-photonic quantum repeaters}.
\newblock {\em Nature Communications}, 6:1--7, 2015.

\bibitem{Munro2015}
William~J. Munro, Koji Azuma, Kiyoshi Tamaki, and Kae Nemoto.
\newblock {Inside Quantum Repeaters}.
\newblock {\em IEEE Journal of Selected Topics in Quantum Electronics}, 21(3),
  2015.

\bibitem{Humphreys2017}
Peter~C. Humphreys, Norbert Kalb, Jaco P.~J. Morits, Raymond~N. Schouten,
  Raymond F.~L. Vermeulen, Daniel.~J. Twitchen, Matthew Markham, and Ronald
  Hanson.
\newblock {Deterministic delivery of remote entanglement on a quantum network}.
\newblock pages 1--7, 2017.

\bibitem{Inlek2017}
I.~V. Inlek, C.~Crocker, M.~Lichtman, K.~Sosnova, and C.~Monroe.
\newblock {Multispecies Trapped-Ion Node for Quantum Networking}.
\newblock {\em Physical Review Letters}, 118(25):1--5, 2017.

\bibitem{simulaqron}
{SimulaQron}.
\newblock \url{http://www.simulaqron.org}.

\bibitem{2058-9565-3-3-034002}
F~Rozp{\c{e}}dek, K~Goodenough, J~Ribeiro, N~Kalb, V~Caprara Vivoli,
  A~Reiserer, R~Hanson, S~Wehner, and D~Elkouss.
\newblock {Parameter regimes for a single sequential quantum repeater}.
\newblock {\em Quantum Science and Technology}, 3(3):34002, 2018.

\bibitem{Krovi2016}
Hari Krovi, Saikat Guha, Zachary Dutton, Joshua~A. Slater, Christoph Simon, and
  Wolfgang Tittel.
\newblock {Practical quantum repeaters with parametric down-conversion
  sources}.
\newblock {\em Applied Physics B: Lasers and Optics}, 122(3):1--8, 2016.

\bibitem{Muralidharan2014}
Sreraman Muralidharan, Jungsang Kim, Norbert L{\"{u}}tkenhaus, Mikhail~D.
  Lukin, and Liang Jiang.
\newblock {Ultrafast and fault-tolerant quantum communication across long
  distances}.
\newblock {\em Physical Review Letters}, 112(25):1--6, 2014.

\bibitem{netsquid}
{NetSQUID: Currently in preparation}.

\bibitem{forwardCitation}
S.~Wehner, A.~Dahlberg, E.~van Zwet, and R.~Hanson.
\newblock A link layer protocol for quantum networks.
\newblock 2018.
\newblock In preparation.

\bibitem{Steiger2016}
Damian~S. Steiger, Thomas H{\"{a}}ner, and Matthias Troyer.
\newblock {ProjectQ: An Open Source Software Framework for Quantum Computing}.
\newblock dec 2016.

\bibitem{Wecker2014}
Dave Wecker and Krysta~M. Svore.
\newblock {LIQUi|{\textgreater}: A Software Design Architecture and
  Domain-Specific Language for Quantum Computing}.
\newblock 2014.

\bibitem{rigetti}
{Forest by Rigetti}.
\newblock \url{https://www.rigetti.com/forest}.

\bibitem{Khammassi2017}
N.~Khammassi, I.~Ashraf, X.~Fu, C.G. Almudever, and K.~Bertels.
\newblock {QX: A high-performance quantum computer simulation platform}.
\newblock {\em Design, Automation {\&} Test in Europe Conference {\&}
  Exhibition (DATE), 2017}, pages 464--469, 2017.

\bibitem{qexperience}
{IBM Q Experience}.
\newblock \url{https://quantumexperience.ng.bluemix.net/qx/experience}.

\bibitem{microsoftQDK}
{Microsoft Quantum Development Kit}.
\newblock \url{https://www.microsoft.com/en-us/quantum/development-kit}.

\bibitem{Johansson2012}
J.~R. Johansson, P.~D. Nation, and Franco Nori.
\newblock {QuTiP: An open-source Python framework for the dynamics of open
  quantum systems}.
\newblock {\em Computer Physics Communications}, 183(8):1760--1772, 2012.

\bibitem{python}
{Python}.
\newblock \url{https://www.python.org/}.

\bibitem{twisted}
{Twisted}.
\newblock \url{https://twistedmatrix.com/trac/}.

\bibitem{Nielsen2010}
Michael~A. Nielsen and Isaac~L. Chuang.
\newblock {\em {Quantum Computation and Quantum Information}}.
\newblock Cambridge University Press, Cambridge, 2010.

\bibitem{mooc}
{MOOC: Quantum Cryptography on}.
\newblock \url{https://www.edx.org}.

\end{thebibliography}
\end{document}